\documentclass[twocolumn]{aastex63}
\usepackage{graphicx}
\usepackage{amsmath}
\usepackage{apjfonts,natbib}
\usepackage{appendix}
\usepackage{lineno}

\usepackage{amsmath}

\newcommand{\oiii}{[O\,{\footnotesize III]}}
\newcommand  \OIII  {\ifmmode \left[{\rm O}\,{\textsc III}\right]\,\lambda5007 \else [O\,{\sc iii}]\,$\lambda5007$\fi}
\newcommand{\Mgiia}{Mg{\sc~ii}\,$\lambda$2796}

\newcommand{\Mgiiab}{Mg{\sc~ii}\,$\lambda\lambda$2796,2803}
\newcommand{\feii}{Fe\,{\footnotesize II}}
\newcommand  \siiv  {\ifmmode {\rm Si}\, {\sc iv}\ \else Si\,{\sc iv}\fi}
\newcommand  \SIIV  {\ifmmode {\rm Si}\,{\sc iv}\,\lambda1399 \else Si\,{\sc iv}\,$\lambda1399$\fi}
\newcommand  \civ  {\ifmmode {\rm C}\, {\textsc IV}\ \else C\,{\sc IV}\fi}
\newcommand  \CIV  {\ifmmode {\rm C}\,{\sc iv}\,\lambda1549 \else C\,{\sc iv}\,$\lambda1549$\fi}
\newcommand  \NV  {\ifmmode {\rm N}\,{\sc v}\,\lambda1240 \else N\,{\sc v}\,$\lambda1240$\fi}
\newcommand  \nv  {\ifmmode {\rm N}\,{\sc v}\ \else N\,{\sc v}\fi}
\newcommand  \cv  {\ifmmode {\rm C}\,{\sc v}\ \else C\,{\sc v}\fi}
\newcommand  \LyA  {\ifmmode {\rm Ly}\,{\sc $\alpha$}\,\lambda1216 \else Ly\,{\sc $\alpha$}\,$\lambda1216$\fi}
\newcommand  \lya {\ifmmode {\rm Ly}\,{\sc $\alpha$}\ \else Ly\,{\sc $\alpha$}\fi}

\newcommand \Hbeta {\ifmmode {\rm H}\beta \else H$\beta$\fi}
\newcommand \hb    {\ifmmode {\rm H}\beta \else H$\beta$\fi}
\newcommand  \mgii  {\ifmmode {\rm Mg}{\textsc{ii}} \else Mg\,{\sc ii}\fi}
\newcommand  \NEV   {\ifmmode \left[{\rm Ne}\,{\textsc v}\right]\,\lambda3426 \else [Ne\,{\sc v}]\,$\lambda3426$ \fi}
\newcommand  \nev  {\ifmmode \left[{\rm Ne}\,{\textsc v}\right] \else [Ne\,{\sc v}]\fi}
\newcommand{\kms}{\ifmmode {\rm km\,s}^{-1} \else km\,s$^{-1}$ \fi}
\newcommand{\cc}{\hbox{cm$^{-3}$}}

\newcommand{\mbh}{\ifmmode M_{\rm BH} \else $M_{\rm BH}$\fi}
\newcommand{\lmbh}{\ifmmode \log\left(\mbh/\Msun\right) \else $\log\left(\mbh/\Msun\right)$\fi}
\newcommand{\lbol} {\ifmmode L_{\rm bol} \else $L_{\rm bol}$\fi}
\newcommand{\ergs}{\ifmmode {\rm ergs\,{\rm s}^{-1} \else ergs\,s$^{-1}$\fi}}

\def\dif{\mathop{}\hphantom{\mskip-\thinmuskip}\mathrm{d}}%
\let\daccent\d
\gdef\d{\ifmmode\dif\else\expandafter\daccent\fi}

\shorttitle{Inflow transition into outflow}
\shortauthors{He et al}
\begin{document}


\title{The Transition from Galaxy-wide Gas Inflow to Outflow in Quasar Host Galaxies}

\correspondingauthor{Zhicheng He}
	\email{zcho@ustc.edu.cn}
\author[0000-0003-3667-1060]{Zhicheng He}
\affiliation{CAS Key Laboratory for Research in Galaxies and Cosmology, Department of Astronomy, University of Science and Technology of China, Hefei, Anhui 230026, China}

\correspondingauthor{Zhifu Chen}
	\email{chenzhifu@gxmzu.edu.cn}
\author{Zhifu Chen}
\affiliation{Department of Physics, Guangxi Minzu University, Nanning 530006, China}

\author{Guilin Liu}
\affiliation{CAS Key Laboratory for Research in Galaxies and Cosmology, Department of Astronomy, University of Science and Technology of China, Hefei, Anhui 230026, China}

\author{Tinggui Wang}
\affiliation{CAS Key Laboratory for Research in Galaxies and Cosmology, Department of Astronomy, University of Science and Technology of China, Hefei, Anhui 230026, China}

\author{Luis C. Ho}
\affiliation{Kavli Institute for Astronomy and Astrophysics, Peking University, Beijing 100871, China}

\author{Junxian Wang}
\affiliation{CAS Key Laboratory for Research in Galaxies and Cosmology, Department of Astronomy, University of Science and Technology of China, Hefei, Anhui 230026, China}

\author{Weihao Bian}
\affiliation{School of Physics, Nanjing Normal University, Nanjing 210046, China}

\author{Zheng Cai}
\affiliation{Department of Astronomy, Tsinghua University, Beijing 100084, China}

\author{Guobin Mou}
\affiliation{School of Physics, Nanjing Normal University, Nanjing 210046, China}

\author{Qiusheng Gu}
\affiliation{School of Astronomy and Space Science, Nanjing University, Nanjing 210093, China}

\author{Zhiwen Wang}
\affiliation{Department of Physics, Guangxi Minzu University, Nanning 530006, China}

\begin{abstract}
Galactic-wide outflows driven by active galactic nuclei (AGNs) is a routinely invoked feedback mechanism in galaxy evolution models. Hitherto, the interplay among the interstellar gas on galactic scales, the propagation of AGN outflows and the fundamental AGN parameters during evolution remains elusive. Powerful nuclear outflows are found to favorably exist at early AGN stages usually associated with high accretion rates and weak narrow emission lines. In a sample of quasars emitting \mgii\ narrow absorption lines (NALs) from the Sloan Digital Sky Survey, we discover an unprecedented phenomenon where galaxy-scale inflow-dominated transforming into outflow-dominated gas accompanied by an increasing strength of the narrow \oiii\ line, at a confidence level of 6.7$\sigma$. The fact that nuclear outflows diminish while galaxy-wide outflows intensifies as AGNs evolve implies that early-stage outflows interact with interstellar medium on galactic scales and trigger the gradual transformation into galaxy-wide outflows, providing observational links to the hypothetical multi-stage propagation of AGN outflows that globally regulates galaxy evolution.
\end{abstract}



\section{Introduction}
AGNs or quasar-driven outflows are considered to play a significant role in shaping the global properties of the host galaxy \citep{Silk1998,2000MNRAS.311..576K,2005Natur.433..604D,2006ApJS..163....1H,fabian2012,zubovas2012,2014ARA&A..52..589H,chen2022}. The outflows span a wide range of scales, ranging from sub-parsec \citep{murray1995,proga2000,zhang2015, williams2017, hamann2018} and a few to tens of parsecs \citep{borguet2012,lucy2014,he2017,he2019,he2022} to scales surpassing kiloparsec (galactic-scale outflow)\citep{liu2013,liu2015,ishibashi2015,zakamska2016,chen2018,shaban2022,shen2023}. Galactic-scale outflows are thought to
be the consequence of the interaction between the AGN (comprising radiation, high-velocity outflows or jets) and the ISM,
ultimately being the fuel for the metal-enriched circumgalactic medium (CGM) \citep{becker2019} and perhaps even intergalactic medium (IGM) \citep{wu2021}. Consequently, such galaxy-wide outflows prove to be a crucial component in comprehending the AGN feedback process. Furthermore, the recycling gas \citep{gaspari2017,tumlinson2017,zhang2023}, consisted of AGN inflow (feeding) and outflow (feedback), regulates the coevolution of supermassive black holes (SMBHs) and galaxies. Yet, certain fundamental correlations between the inflow and outflow, such as the variation in the detection ratio of outflow to inflow throughout various AGN evolutionary stages or its relationship with the fundamental parameters of AGN, have not been adequately explored. If the ISM or galaxy-scale inflow gas interacts with the AGN radiation and nuclear outflow, leading to the formation of a galactic-scale outflow, a natural consequence is the mutual transformation of the inflow and outflow processes. However, it is impossible to observe such a transformation at the galactic-scale outflow for a single galaxy within a human lifespan. Fortunately, the sample statistics can offer these insights, by analysing this transformation phenomenon through an extensive sample of AGNs instead of individual sources. In this study, using a large sample, we explore the transition of galactic-scale inflow into outflow, and its correlation with the fundamental parameters of AGNs, aiming to reveal their feedback influence on the host galaxies.

Researches (e.g., \citealt{boroson1992, boroson2002, shen2014}) on the main sequence of AGNs have revealed that the diversity of AGNs could be explained through two key factors, accretion and orientation (see Figure 1 of ref. \citealt{shen2014}). In Figure 1 of ref. \cite{shen2014}, the systematic trend of decreasing strength of narrow \OIII\ emission is observable with increasing strength of \feii\ relative to the broad \hb\ ($R_{\rm \feii}=EW_{\rm \feii}/EW_{\rm \hb}$). This is one of the characteristics of the 'Eigenvector 1' (EV1) sequence. The EV1 is a principal component composed of the correlation coefficients of many variables, which has long been considered to be driven by the Eddington ratio. Among all variables, the correlation coefficient of $R_{\rm \feii}$ is almost the largest. In addition, in AGNs with high Eddington ratios, $R_{\rm \feii}$ is generally strong, e.g., the Narrow Line Seyfert 1 (NLS1s) galaxies \citep{grupe1999,mathur2000,grupe2004, komossa2018}. Therefore, $R_{\rm \feii}$ is considered an indicator to the Eddington ratio. What does the changing trend of the narrow \oiii\ strength in AGNs represent? AGNs are believed to be triggered, potentially, by the merger or interaction of galaxies \citep{sanders1990,veilleux2009,hopkins2016}. In the early stage of AGN evolution, galaxies typically contain dense gas and a significant amount of dust. As AGNs evolve, the ionization zone resulting from the UV radiation of the AGN gradually expands.
Through the proximity effect\citep{carswell1982,murdoch1986,bajtlik1988}, as reported in ref. \cite{zheng2020}, it has been found that older AGNs (corresponding to larger proximity zones) exhibit greater equivalent widths (EWs) of narrow emission lines compared to younger AGNs (corresponding to smaller proximity zones), based on a sample of approximately 2600 SDSS quasar spectra. Additionally, considered as a type of young AGNs, the NLS1s also exhibit the characteristics of weak \oiii\ strength \citep{grupe1999,mathur2000,grupe2004, komossa2018}.
Here, we redraw the Figure 1 of ref. \cite{shen2014} with the catalog of Quasar Properties from SDSS Data Release 16 (DR16Q)\citep{wu2022} (see Appendix section \ref{sec:a1} for sample description).
As shown in panels (a) and (b) of Fig. \ref{fig:a1}, the systematic trend of decreasing $R_{\rm \feii}$ with increasing \oiii\ strength indicates that the accretion rate becomes weaker as AGN evolve. In panels (c) and (d) of Fig. \ref{fig:a1}, a moderate negative correlation between \oiii\ strength and Eddington ratio is observed.
In this study, we explore the factors behind the increase in \oiii\ emission with AGN ages, while simultaneously utilizing \oiii\ emission as an age indicator of AGNs to examine the
evolution of high-speed outflow from nuclear region and galactic-scale outflows.
Considering the higher correlation coefficient between the luminosity ratio of \oiii\ to 3000\AA\ ($L_{\rm \oiii}/L_{3000}$) and Eddington ratio (Spearman correlation coefficient, $r=-0.3$) compared to the EW of \oiii\ and Eddington ratio ($r=-0.21$), here we utilize the $L_{\rm \oiii}/L_{3000}$ as indicators of AGN age.

\section{Results}

\begin{figure*}
\centering
\includegraphics[width=14.0cm]{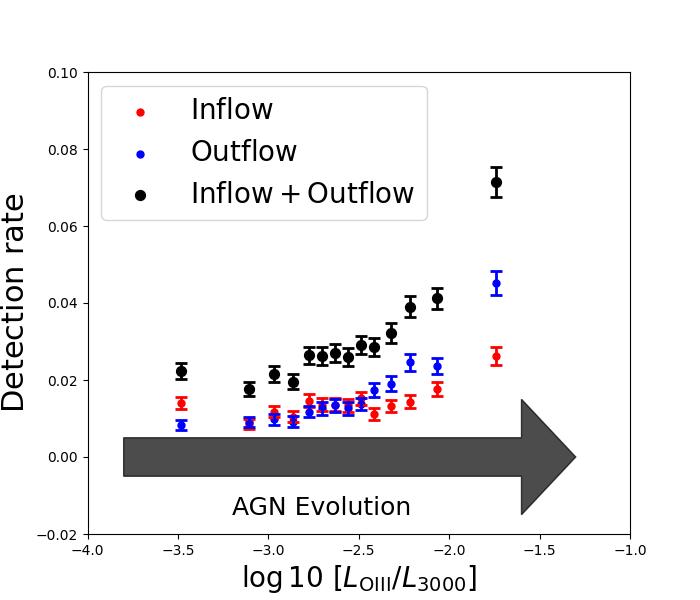}
\caption{The detection rates of outflow and inflow using \mgii\ NALs across varying \OIII\ to 3000\AA\ luminosity ratios which considered the indicators of AGN evolutionary directions. Both the detection rates of inflow (redshift NALs) and outflow (blueshift NALs) are increased with the value of $L_{\rm \oiii}/L_{3000}$. More noteworthy is that, the rate of increase in outflow fraction is faster than that of inflow. }
\label{fig1}
\end{figure*}

Absorption line spectroscopy has been extensively used for more than four decades to study the distribution of gas around galaxies, including broad absorption lines (BALs) and NALs. The BALs often have a typical line width greater than 2000 \kms\citep{Weymann1991}, and the NALs typically have velocity widths being less than a few hundred \kms. It is widely believed that the BALs are typically formed by the high-speed outflow of AGN nuclear regions, including accretion disk winds\citep{murray1995,proga2000} and dusty torus winds
\citep{konigl1994,scoville1995,gallagher2015,he2022}. The NALs usually have a complex origination. They can be produced by the materials physically associated with quasar systems and the intervening galaxies far beyond quasar systems. It is believed that the NALs would be mainly originated in quasar host galaxies when their velocity offsets ($\upsilon_r$) less than 1000 km/s relative to quasar system redshifts \citep{2018ApJS..235...11C}. However, there is no strict boundary between BAL and NAL. The intrinsic NAL of AGN system may correspond to a position closer to the nucleus region as the velocity increases.

\begin{figure*}
\centering
\includegraphics[width=16.0cm]{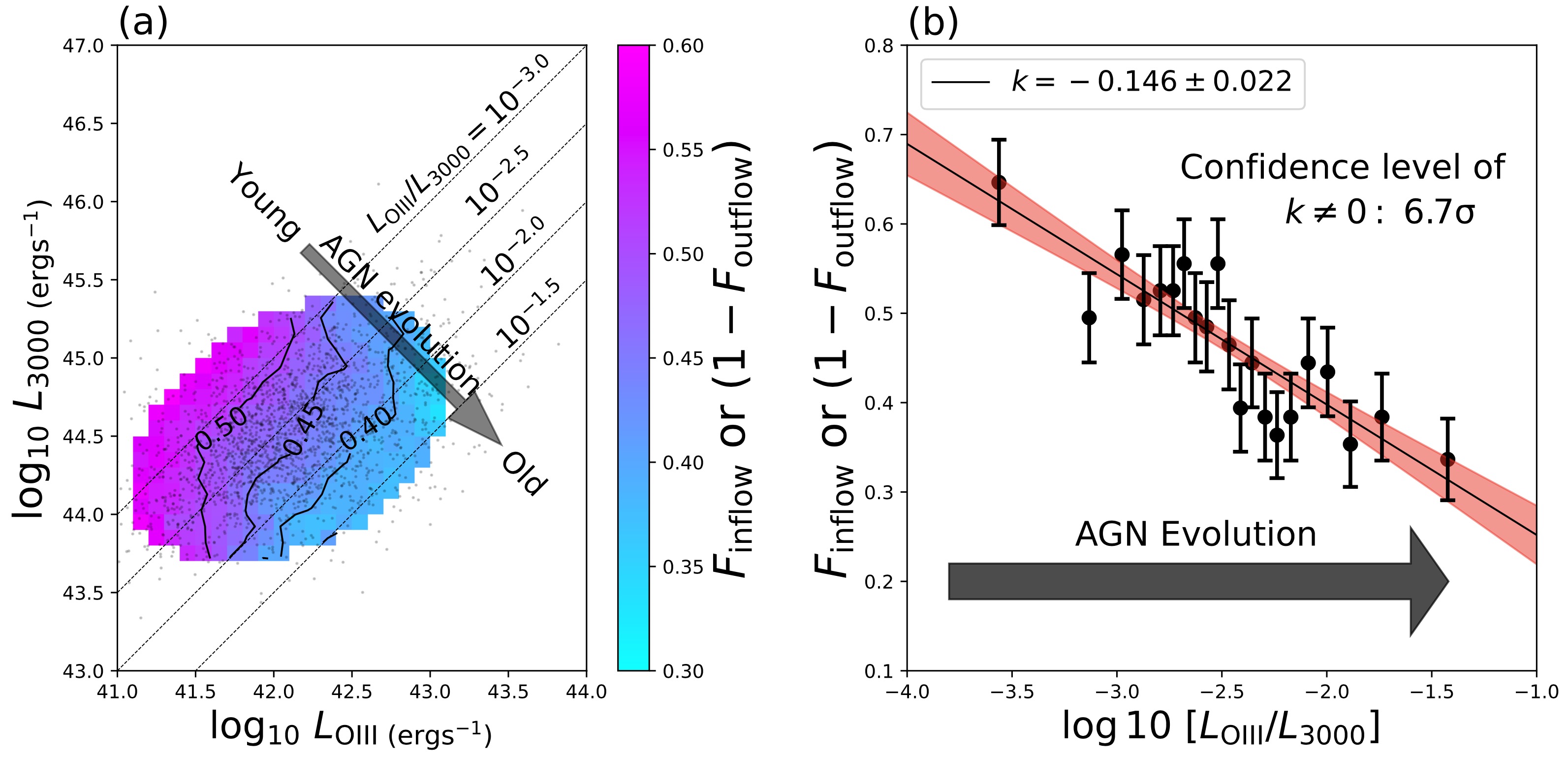}
\caption{The transformation phenomenon from galaxy-wide gas inflow and outflow. Panel (a), the black points are the $L_{3000}$ versus $L_{\rm \oiii}$ for the 2108 \mgii\ NALs from SDSS. The black arrow indicates the direction in which the value of $L_{\rm \oiii}$/$L_{3000}$ or $EW_{\rm \oiii}$ increases, which represents the direction of AGN evolution. The $F_{\rm inflow}$ decreases as the ratio $L_{\rm \oiii}$/$L_{3000}$ increases. Panel (b), the 2108 AGNs are divided into 21 bins. The first 20 bins contain 100 objects each, while the last bin contains 108 objects. The slope of the best linear fit is $k=-0.146\pm 0.022$. The red shadow is 1 $\sigma$ confidence region. The confidence level of the slope deviating from $k=0$ is $6.7\sigma$. Our results reveal that as AGN evolves, more and more inflows gradually transform into outflow. }
\label{fig2}
\end{figure*}

In this study, we adopt \Mgiiab\ NALs with line widths less than 400 \kms\ and velocity offsets $\upsilon_r<500$ \kms\ (also take 750 \kms\ and 1000 \kms\ as a comparison) to trace the low velocity gas in AGN host galaxy. Here we have collected 2108 \mgii\ NALs (see Appendix section \ref{sec:a2} for \mgii\ NALs certification) from a sample of 69,536 quasars obtainted from the SDSS DR16Q\citep{2020ApJS..250....8L}. Among these NALs, we identified 977 cases of redshifted (inflows) and 1,131 cases of blueshifted (outflows) \mgii\ NALs. Examples of the redshifted and blueshifted \mgii\ NALs are shown in Fig. \ref{fig:a2}, respectively. We define the detection rates of inflow and outflow as: $D_{\rm inflow}=N_{\rm inflow}/N_{\rm sample}$ and $D_{\rm outflow}=N_{\rm outflow}/N_{\rm sample}$, respectively, where $N_{\rm inflow}$ represents the total number of the redshifted \mgii\ NALs, $N_{\rm outflow}$ indicates the  total number of the redshifted \mgii\ NALs, and $N_{\rm sample}$ is the number of the total sample. Fig. \ref{fig1} clearly shows that both the detection rates of the inflows and outflows are increased with the value of $L_{\rm \oiii}/L_{3000}$. More noteworthy is that, the rate of increase in outflow detection $D_{\rm outflow}$ is faster than that of inflow detection $D_{\rm inflow}$. In order to carefully analyze this phenomenon, we plot the $L_{3000}$ and $L_{\rm \oiii}$ for the 2108 \mgii\ NALs in panel (a) of Fig. \ref{fig2}. The fraction of inflows ($F_{\rm inflow}$) is defined as: $F_{\rm inflow}=(1-F_{\rm outflow})=N_{\rm inflow}/(N_{\rm inflow}+N_{\rm outflow})$. To calculate $F_{\rm inflow}$ values, we consider only the regions with at least 100 objects within a radius of 0.3 dex, using a step size of 0.1 dex in the $L_{3000}$ versus $L_{\rm \oiii}$ planes. As illustrated in panel (a) of Fig. \ref{fig2}, a clear and unprecedented evolutionary trend is observed for the $F_{\rm inflow}$ across the $L_{3000}$ versus $L_{\rm \oiii}$ planes. The $F_{\rm inflow}$ decreases as the ratio $L_{\rm \oiii}$/$L_{3000}$ increases, indicating a progressive transition from inflow-dominated to an outflow-dominated population. To quantitatively investigate the correlation between $F_{\rm inflow}$ and $L_{\rm \oiii}/L_{3000}$, we performed a sorting of the 2108 AGNs based on the value of $L_{\rm \oiii}/L_{3000}$ and divided them into 21 bins. The first 20 bins comprise 100 objects each, and the last bin contains 108 objects. The relationship between $F_{\rm inflow}$ and $L_{\rm \oiii}/L_{3000}$ for these 21 bins is graphically presented in panel (b) of Fig. \ref{fig2}. Utilizing  the KMPFIT method, we obtained the best linear fit for $F_{\rm inflow}$ as $(-0.146\pm 0.022)\log_{10}(L_{\rm \oiii}$/$L_{3000})+(0.106\pm 0.054)$. The confidence level of the fitting slope, $k=-0.146\pm 0.022$, deviates significantly from $k=0$ at $6.7\sigma$. Accordingly, the probability of no correlation between $F_{\rm inflow}$ and $L_{\rm \oiii}/L_{3000}$ is remarkably low, with a value of $p_{\rm null}=2.5\times 10^{-11}$. This unprecedented result provides conclusive evidence of galactic-scale inflows gradually transitioning into outflow-dominated population during AGN evolution.

\begin{figure*}
\centering
\includegraphics[width=16.cm]{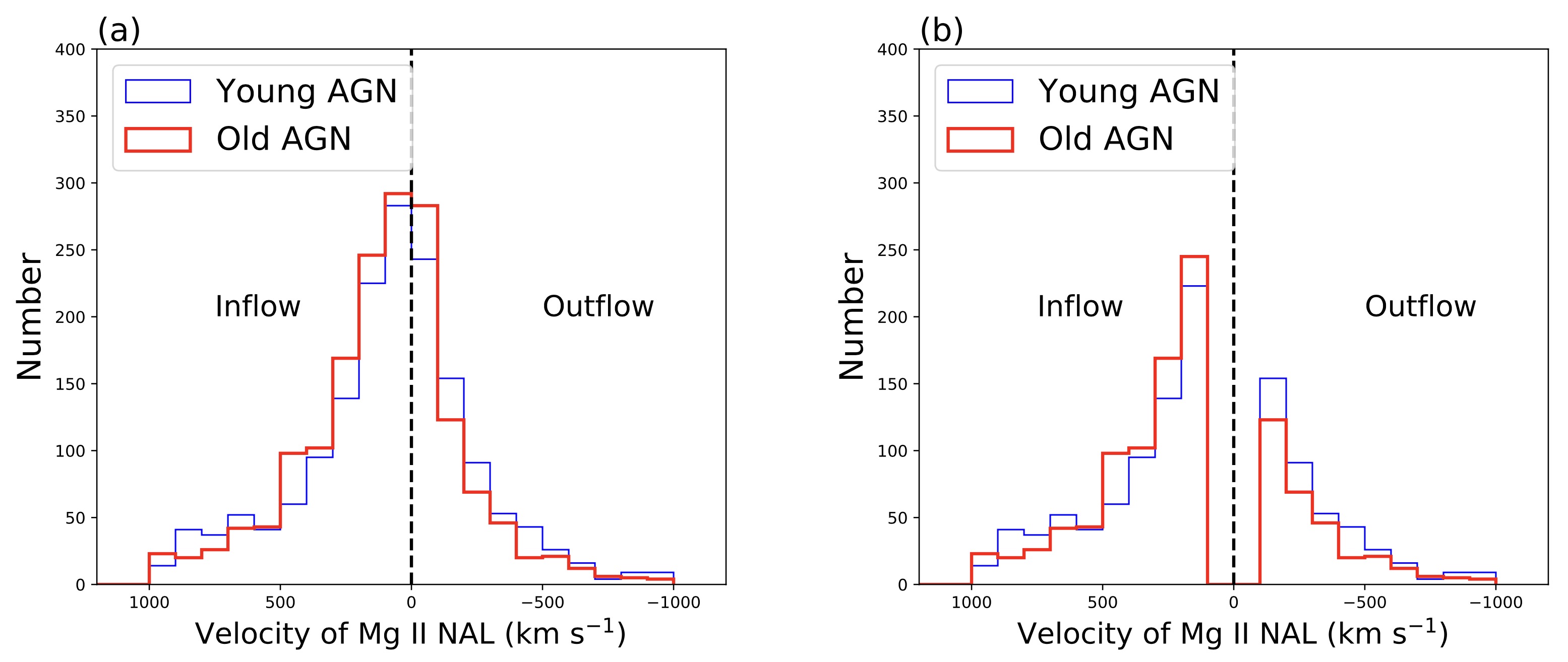}
\caption{Comparison of NALs velocity distributions between Young and Old AGNs. Panel (a) is for  the whole sample and panel (b) has excluded those with NAL velocities less than 100 \kms. Positive velocity represents inflow, while negative velocity represents outflow. Panel (a): compared with younger AGNs, the NAL average velocity in older AGNs is more blue shifted (from -77 \kms\ to -101 \kms). Like the fraction of inflow, the velocity of inflow is also decreasing (223 \kms\ to 190 \kms). Interestingly, although the fraction of outflow is increasing, the absolute value of the outflow average velocity is decreasing (from 277 \kms\ to 264 \kms). Panel (b): the evolution of NAL velocity shows the same trend as panel (a). The average NAL velocity in young and old AGNs are -111 \kms\ and -154 \kms. The average inflow velocity in young and old AGNs are 324 \kms\ and 316 \kms. The absolute value of the outflow average velocity in young and old AGNs are 367 \kms\ and 346 \kms.}
\label{fig3}
\end{figure*}

We further divide the sample into two groups based on \oiii\ : young (i.e., $L_{\rm \oiii}/L_{3000}$ smaller than the mean value $10^{-2.4}$) and old ($L_{\rm \oiii}/L_{3000}>10^{-2.4}$), and then compare the velocity distributions of the two groups. As shown in Fig. \ref{fig3}, panel (a) is for the whole sample and panel (b) has excluded those with NAL velocities less than 100 \kms. Positive velocity represents inflow, while negative velocity represents outflow. Panel (a) of Fig. \ref{fig3} compared with younger AGNs, the NAL average velocity in older AGNs is more blue-shifted (from -77 \kms to -101 \kms). The probability that there is no difference in the average value between the two through T-test is $p_{\rm null}=0.03$. Like the fraction of inflow, the velocity of inflow is also decreasing (223 \kms\ to 190 \kms, with $p_{\rm null}=0.007$). Interestingly, although the fraction of outflow is increasing, the absolute value of the outflow average velocity is decreasing (from 277 \kms\ to 264 \kms, with $p_{\rm null}=0.2$). This result suggests that the high-velocity outflow from the center triggers more low-speed outflow. As a result, the average velocity decreases, but the fraction of outflow increases. As shown in panel (b) of  Fig. \ref{fig3}: the evolution of NAL velocity shows the same trend as panel (a). The NAL average velocity in young and old AGNs are -111 \kms and -154 \kms, with $p_{\rm null}=0.01$. The inflow average velocity in young and old AGNs are 324 \kms and 316 \kms, with
$p_{\rm null}=0.64$. The absolute value of the outflow average velocity in young and old AGNs are 367 \kms and 346 \kms, with
$p_{\rm null}=0.06$. In the future, larger samples are expected to increase the confidence level of differences in velocity distribution.

\section{Discussions}

\begin{figure*}
\centering
\includegraphics[width=14.0cm]{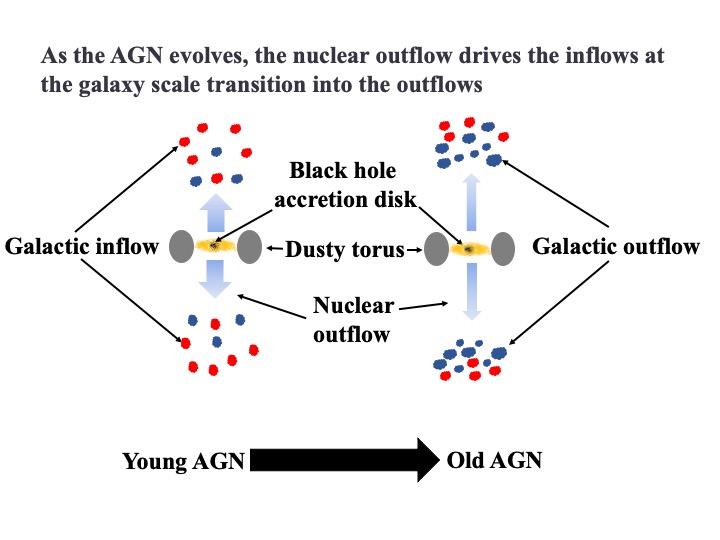}
\caption{A schematic diagram of the evolution of high-speed outflows from the nuclear region and low-speed outflows in the galaxy scale. In the early stages of AGN evolution, high accretion rate from AGN drives strong outflow and at this moment, the materials at the galaxy scale is mainly inflow. As AGN evolves, the accretion rate gradually weakens, resulting in a decrease in the intensity of nuclear outflow. The previous nuclear outflow has reached the galactic scale and driven the cold cloud to gradually transition from inflow to outflow. The fragmentation or expansion of clouds due to the hit of nuclear outflows leads to an increase in both [O III] emissions and the detection rate of Mg II NALs.}
\label{fig4}
\end{figure*}

Interestingly, on the contrary, many studies \citep{baskin2005,ganguly2007,komossa2008,zhang2014,schmidt2018} on the broad absorption lines (BALs) and the asymmetry or blue-shift of emission lines (e.g., \civ, \oiii) reveal that the central strong outflow favorably exists in early stage of AGNs with high accretion rates. To sum up, we are currently faced with four questions awaiting our solutions: \textbf{1, What is the reason behind the increase in \oiii\ emissions as AGN ages? 2, Why does the detection rate of \mgii\ NALs increase with \oiii\ emission? 3, Why does the outflow (blueshift \mgii\ NALs) detection rate increase faster than that of the inflow (redshift \mgii\ NALs)? 4, Why do high-speed outflows from the nuclear region tend to occur during the early stages of AGNs, while galaxy-scale outflows are more prevalent in the late stages?} Here, we propose the following scenario to answer the above questions simultaneously. As shown in Fig. \ref{fig4}, in the early stages of AGN evolution, the high accretion rate of AGN drives strong outflow and at this moment, the materials at the galaxy scale are mainly inflows. As AGN evolves, the accretion rate gradually weakens, resulting in a decrease in the strength of nuclear outflow. The previous nuclear outflow has reached the galactic scale and driven the ISM to gradually transition from an inflow-dominated to an outflow-dominated population. Through hydrodynamic simulations, as referenced in ref. \cite{hopkins2010}, it was discovered that when a cold, denser cloud is struck by a diffuse outflow, it can either fragment or expand. The Cloudy \citep{ferland2017} simulation (see Appendix section \ref{sec:a3} and Fig. \ref{fig:a4} for details) demonstrates that the \oiii\ emission predominantly concentrates on the surface of the cloud that faces the central radiation source. For a gas cloud with volume density $n_{\rm H}=10^4\cc$, side length $l$ =1 pc and distance $R$=1 kpc from the central radiation source (with a typical Ionizing photon emissivity $Q_{\rm H}=10^{55}\rm s^{-1}$). As shown in panel (a) of Fig. \ref{fig:a4}, the \oiii\ emission predominantly concentrates on the surface within depth of $d=10^{15.5}$ cm ($\sim$ 0.001 pc). So, when the cloud cube is fragmented into 10 pieces, the \oiii\ emission doubles (panel (c)) due to the larger surface area. Alternatively, when the cloud expands 10 times ($n_{\rm H}$ decreases from $10^4\cc$ to $10^3\cc$ ), \oiii\ emissions increase by more than 20 times. Hence, the fragmentation or expansion of clouds caused by the nuclear outflow has a dual effect. It not only results in an increased surface area facing the central source (i.e., an expanded solid angle) but also leads to elevated \oiii\ emissions. Consequently, this process can account for the increase in both \oiii\ emissions and the detection rate of \mgii\ NALs as AGN ages. In summary, our scenario simultaneously addresses the four aforementioned issues. The fact that nuclear outflows diminish while galaxy-wide outflows intensifies as AGNs evolve implies that early-stage outflows interact with interstellar medium on galactic scales and trigger the gradual transition into galaxy-wide outflows, providing observational links to the hypothetical multi-stage propagation of AGN outflows that globally regulates galaxy evolution.

It's should note that the vast majority of sources in our sample are Type 1 QSOs with $FWHM_{\hb} \ge 1500\kms$. Therefore, the variation in the solid angle of the ionization zone cannot account for the detection rate of \mgii\ NALs. Furthermore, some studies \citep{ricci2017,ricci2022,zhuang2018} have indicated that as the accretion rate decreases, the angle of the ionization region also appears to decrease, making it increasingly challenging to account for the rise in \oiii\ emission and the detection rate of \mgii\ NALs during AGN evolution. We also considered whether the increase in the fraction of outflow  with increasing narrow \oiii\ emission was caused by the inclination effect. When the inclination angle increases and approaches the dust torus, the gas can be accelerated more effectively, resulting in a higher fraction of outflow. Meanwhile, due to the influence of dust extinction, the continuum will weaken, leading to an increase in the relative intensity of \oiii\ emission lines. If the inclination effect takes effect, the larger the inclination angle, the more dust there will be, resulting in a stronger narrow \oiii\ emission and a shorter length of the ionization zone measured by the proximity effect. However, the actual observation result \citep{zheng2020} is that the longer the ionization zone, the stronger the narrow \oiii\ emission. Therefore, the inclination effect can be excluded. Furthermore, we investigated the influence of quasar redshift on the relationship between the fraction of inflow and AGN evolution (see Appendix section \ref{sec:a4} for details). The sample was divided into two groups based on the median redshift value ($z_{\text{median}} = 0.72$). Fig. \ref{fig:a5} illustrates that the confidence level of the $F_{\text{inflow}}$ decreasing with increasing $L_{\text{\oiii}}/L_{3000}$ is higher for the quasars with $z<0.72$ compared to the quasars with $z>0.72$. This discrepancy may be attributed to the presence of more foreground-inserted non-intrinsic blue-shifted absorption lines (misidentified as outflows) in the quasars with $z>0.72$, which can impact the results. The outflow fractions for the quasars with $z<0.72$ and $z>0.72$ are 545/1054 and 586/1054, respectively, providing further support for the higher fraction of blue-shifted absorption lines in the quasars with $z>0.72$. In view of this, the absorption lines caused by the foreground-inserted materials will contaminate the signal of the phenomenon of galaxy-scale inflow-dominated transition into outflow-dominated gas. The confidence level of the intrinsic phenomenon of galaxy-scale inflow transforming into outflow should be higher than the measured value.
Finally, we examined the influence of the outflow velocity threshold on the results (see Appendix section \ref{sec:a4} and Fig. \ref{fig:a6} for details).
We found that the influence of different thresholds (500 \kms, 750 \kms, and 1000 \kms) can be ignored.

\section{Conclusions}

AGNs or quasar-driven outflows are considered to play a significant role in shaping the global properties of the host galaxy.
If the high-speed outflow of the nuclear region affects the star formation of the entire galaxy, it will inevitably first affect the distribution or motion characteristics of gas at the galaxy scale.
Therefore, in the evolution process of active galaxies, the detection rates of gas inflow and outflow in the galaxy will undergo corresponding evolution.
In this work, we analyzed the evolution of \mgii\ NAL detection rate using SDSS DR16Q samples and \oiii\ narrow emission line as an indicator for AGN age.
We found that the fraction of NAL inflow (outflow) gradually decreases (increases) with the evolution of AGN, reaching a confidence level of 6.7$\sigma$.
This means that during the evolution of AGN, inflow gradually transforms into outflow.
In addition, although the fraction of outflows has increased with evolution, the average outflow speed is decreasing.
This indicates that the low-speed large-scale outflow is increasing compared to the high-speed central outflow.
Many previous studies have found that as AGN evolves, the central accretion rate weakens, and the high-speed outflow from the center also weakens.
Based on these studies and our work, we have summarized the following evolutionary scenario: in the early stages of AGN evolution, a high accretion rate of AGN drives
strong nuclear outflow and at this moment, the materials at the galaxy scale are mainly inflows. As AGN evolves, the accretion rate gradually weakens,
resulting in a decrease in the strength of nuclear outflow. The previous nuclear outflow has reached the galactic scale and driven the ISM to transition progressively from an
inflow-dominated to an outflow-dominated population.

\acknowledgments
Z. C. He is supported by the National Natural Science Foundation of China (nos. 12222304, 12192220, and 12192221). Z. F. Chen is supported by the Guangxi Natural Science Foundation (2024GXNSFDA010069), the National Natural Science Foundation of China (12073007), and the Scientific Research Project of Guangxi University for Nationalities (2018KJQD01).
We acknowledge the research grants from the China Manned Space Project (No. CMS-CSST-2021-A06 and No. CMSCSST-2021-A07), the National Natural Science
Foundation of China (No. 11421303), the Fundamental Research Funds for the Central Universities (No. WK3440000005), and the K. C. Wong Education Foundation.
L. C. Ho was supported by the National Science Foundation of China (11721303, 11991052, 12011540375, 12233001), the National Key R\&D Program of
China (2022YFF0503401), and the China Manned Space Project (CMS-CSST-2021-A04, CMS-CSST-2021-A06).
J. X. Wang is  supported by the National Science Foundation of China (12033006).

Funding for the Sloan Digital Sky Survey IV has been provided by the Alfred P. Sloan Foundation, the U.S. Department of Energy Office of
Science, and the Participating Institutions.
SDSS-IV acknowledges support and resources from the Center for High Performance Computing  at the University of Utah. The SDSS
website is www.sdss.org.
SDSS-IV is managed by the Astrophysical Research Consortium for the Participating Institutions of the SDSS Collaboration including
the Brazilian Participation Group, the Carnegie Institution for Science, Carnegie Mellon University, Center for
Astrophysics | Harvard \& Smithsonian, the Chilean Participation Group, the French Participation Group, Instituto de Astrof\'isica de
Canarias, The Johns Hopkins University, Kavli Institute for the Physics and Mathematics of the Universe (IPMU) / University of
Tokyo, the Korean Participation Group, Lawrence Berkeley National Laboratory, Leibniz Institut f\"ur Astrophysik
Potsdam (AIP),  Max-Planck-Institut f\"ur Astronomie (MPIA Heidelberg), Max-Planck-Institut f\"ur Astrophysik (MPA Garching),
Max-Planck-Institut f\"ur Extraterrestrische Physik (MPE), National Astronomical Observatories of China, New Mexico State University,
New York University, University of Notre Dame, Observat\'ario Nacional / MCTI, The Ohio State University, Pennsylvania State
University, Shanghai Astronomical Observatory, United Kingdom Participation Group, Universidad Nacional Aut\'onoma
de M\'exico, University of Arizona, University of Colorado Boulder, University of Oxford, University of Portsmouth, University of Utah,
University of Virginia, University of Washington, University of Wisconsin, Vanderbilt University, and Yale University.


\bibliography{ref.bib}
\appendix

\renewcommand{\thesection}{Appendix}

\section{\label{sec:appendix}}

\subsection{Quasar parent sample}\label{sec:a1}
The SDSS collected spectra in the wavelength range $3800-9200$ \AA\ at a resolution of $R\approx2000$ from 2000 to 2008 (SDSS-I/II) \citep{2009ApJS..182..543A}, and in the wavelength range $3600-\rm10~500$ \AA\ at a resolution of $R\approx1300-2500$ from 2008 to 2020 (SDSS-III/IV) \citep{2013AJ....146...32S,2013AJ....145...10D,2016AJ....151...44D}. DR16Q includes 750,414 quasars that are accumulated from SDSS-I to SDSS-IV \citep{2020ApJS..250....8L}, which is the final quasar catalog for the SDSS-IV. We use the spectra from DR16Q to construct a sample of quasars with \OIII\ emission lines and associated \Mgiiab\ NALs. Firstly, we only select the quasars with redshifts $0.35\le z_{em}<1.0$, where the $z_{\rm em}$ is the best redshifts from ref. \cite{wu2022}. This guarantees the spectra from the SDSS that cover the \OIII\ emission lines and associated \mgii\ NALs. Secondly, we retain only those quasars whose \OIII\ emission lines have equivalent widths $EW_{\rm \OIII}\ge4\sigma_{EW_{\rm \OIII}}$, where the $EW_{\rm \OIII}$ can be available in the catalog of quasar properties provided by ref. \cite{wu2022}. The associated \mgii\ NALs are detected in the spectroscopic data around the \mgii\ emission lines of quasars. Hence, thirdly, we include only quasars with median signal-to-noise ratio $S/N > 3$ within the spectral regions of $\pm6000 \kms$ around \mgii\ emission lines, since the spectra with lower $S/N$ are not conducive to detecting \mgii\ NALs \citep{2018ApJS..235...11C}. Based on these criteria, we assemble a sample of 69,536 quasars, which are used to search for associated \mgii\ NALs.

Ref. \cite{shen2014} depicted the distribution of 20,000 quasars in the EV1 plane, revealing a systematic trend of declining narrow \oiii\ strength as the relative strength of \feii\ to the broad \hb\ ($R_{\rm \feii}=EW_{\rm \feii}/EW_{\rm \hb}$) increases. Our sample, as shown in panels (a) and (b) of Fig. A1, exhibits a similar distribution and reproduces this systematic trend. Previous studies have observed that Narrow Line Seyfert 1 galaxies (NLS1s), considered to be in the early stage of AGN evolution \citep{mathur2000, komossa2018}, often display high values of $R_{\rm \feii}$ and weak \oiii\ \citep{grupe2004}. Additionally, utilizing the proximity effect \citep{carswell1982,murdoch1986,bajtlik1988}, ref. \cite{zheng2020} found that older AGNs exhibit stronger narrow emission lines compared to younger AGNs. Furthermore, the detection rates of \mgii\ NALs (Fig. \ref{fig1}) in our sample increase with the strength of narrow \oiii, suggesting that the increasing \oiii\ strength could serve as an indicator of the expanding ionization region as AGNs evolve. Consequently, we adopt the \oiii\ strength as a tracer of AGN evolutionary stages. As depicted in panels (c) and (d) of Fig. \ref{fig:a1}, the inverse correlation between \oiii\ and $R_{\rm \feii}$ implies that the evolution of AGNs is accompanied by a decline in accretion rate.

\subsection{Sample of \mgii\ NALs}\label{sec:a2}
In order to fit a pseudo-continuum (e.g., the red solid-lines shown in Fig. \ref{fig:a2} for each quasar spectra, we invoke a combination of multi-Gaussian functions and cubic spline fitting to capture the underlying continuum and emission lines in an iterative fashion \citep{2005ApJ...628..637N,2018ApJS..235...11C}. The candidates of associated \mgii\ NALs are searched for in the spectroscopic data normalized by the pseudo-continuum fit (e.g., the spectra shown in panels (b) and (c) of Fig. \ref{fig:a2}). This process is mainly controlled by the separation of the \Mgiiab\ doublet. To include the vast majority of redshifted \mgii\ NALs, we search for \mgii\ NALs from the blue wing $\upsilon_r = -1000$ \kms\ until the red wing $\upsilon_r = 6000$ \kms\ of the \mgii\ emission line. We invoke a pair of Gaussian functions to fit each candidate \Mgiiab\ doublet (e.g., the green solid-lines shown in Fig. \ref{fig:a2}), and visually check the fitting results one by one. We retain only \mgii\ absorption lines that satisfy the following criteria: $W_{\rm r}^{\rm \lambda2796}\ge3\sigma_{\rm w}$, $W_{\rm r}^{\rm \lambda2803}\ge2\sigma_{\rm w}$, $W_{\rm r}^{\rm \lambda2796}\ge0.2$ \AA, $W_{\rm r}^{\rm \lambda2803}\ge0.2$ \AA, and $\rm FWHM\le400$ \kms, where $W_r$ is the equivalent width at absorber rest-frame, and $\rm FWHM$ is the full width at half-maximum of the \Mgiia\ absorption line. If one quasar exhibits multiple \mgii\ NALs, we determine whether the NAL is redshifted or blueshifted case based on the average velocity ($\upsilon_r$) of the multiple NALs, where $\upsilon_r$ is computed by
\begin{equation}\label{eq:vr}
  \upsilon_r = c \times \frac{(1 + z_{\rm abs})^2 - (1 + z_{\rm em})^2 }{(1 + z_{\rm abs})^2 + (1 + z_{\rm em})^2 }
\end{equation}
where $c$ is the speed of light, $z_{\rm em}$ is the redshift of the quasar system, and $z_{\rm abs}$ is the redshift of the \mgii\ NALs. Panel (a) of Fig. \ref{fig:a3} shows the velocity distribution of the detected \mgii\ NALs.

After the certification process of the \mgii\ NALs, we obtain 3785 trustworthy \mgii\ NALs from 3747 quasars. To mitigate the potential contamination from non-intrinsic object insertions and high-velocity outflows from the galactic nucleus, we excluded absorption lines with blue-shift velocities greater than 500 \kms. Additionally, to account for uncertainties in redshift measurements, we removed absorption lines in a velocity range of $-100<\upsilon_{\rm r}<100$ \kms. Consequently, we retained 2133 reliable \mgii\ NALs identified from 2108 quasars.

We adopted the improved systemic redshift $\rm Z_{SYS}$ from the catalog of quasar properties provided by ref. \cite{wu2022} as the quasar redshift. The $\rm Z_{SYS}$ is a mean value obtained from multiple individual lines and is considered to be the most accurate redshift measurement according to ref. \cite{wu2022}.  The catalog also provides other quasar properties such as the black hole mass (\mbh, $\rm LOGMBH_HB$), the luminosity at 3000\AA\ ($L_{\rm 3000}$, $\rm LOGL3000$), and the luminosity of the narrow component of the \OIII\ emission line ($\rm L_{\rm \OIII}$, $\rm OIII5007C$), respectively. the velocity distributions  of \mgii\ NALs, quasar redshift, \mbh\ and luminosity at 3000\AA\ are shown in Fig. \ref{fig:a3}.

\subsection{Cloudy simulation for the \OIII\ emission}\label{sec:a3}
In this section, we study the characteristics of \OIII\ emissions from a dense cloud in the NLR of AGN. We adopt the mean spectral energy distribution (SED) of three typical AGN described in ref. \cite{arav2013}: MF87 \citep{mathews1987}, UV-soft \citep{dunn2010} and HE 0238 \citep{arav2013}. The bolometric correction at 3000\AA\ is  $\log \rm BC_{3000} = 0.71$, i.e., $\log \lbol = \log L_{3000}+0.71$. The relationship between the ionizing photon emissivity and bolometric luminosity is: $\log Q_{\rm H}=\log \lbol +9.8$. The median value of quasar luminosity at 3000\AA\ of our sample is $10^{44.5} \ergs $. So, the typical Ionizing photon emissivity of quasar in our sample is $Q_{\rm H}=10^{55}\rm s^{-1}$). For a cube of gas cloud with volume density $n_{\rm H}=10^4~\cc$, side length $l$ =1 pc and distance $R$=1 kpc from the central AGN. The ionization parameter at the surface facing the central AGN is $U=10^{-3.5}$. The outcome of the Cloudy simulation is illustrated in Fig. \ref{fig:a4}. As shown in panel (a), the \oiii\ emission predominantly concentrates on the surface within depth of $d=10^{15.5}$ cm ($\sim$ 0.001 pc) and column density of  $N_{\rm H}=10^{19.5}\rm cm^{-2}$. The total \oiii\ luminosity is $10^{35.9} \ergs$. As shown in panel (c), If the cloud cube is fragmented into 10 identical blocks each with a side length of $l$ =0.46 pc, so the total area facing the central AGN of 10 pieces, will be 2.15 times the original cloud. Then, the \oiii\ emission increases 2.15 times due to the larger surface area.
As shown in panel (b), when the mass of cloud remains unchanged and the volume expands by 10 times ($n_{\rm H}$ decreases from $10^4~\cc$ to $10^3~\cc$ ).
The surface area facing the central AGN in the expanded cloud is 4.6 times that of the original cloud. When combined with the emission coefficient of \oiii\ in panel (b),
the total \oiii\ luminosity increases to $10^{37.3} \ergs$ (a 21-fold increase). Please note that Cube clouds are non-physical and unrealistic. The multiple increase in
\oiii\ luminosity is only a calculated result. Our goal is solely to qualitatively demonstrate that the fragmentation or expansion process of clouds can result in
increased \oiii\ emission.

According to the CLOUDY simulation, for a typical cold flow gas with a temperature of $\rm 10^4 K$, the recombination coefficient of $\rm O^{2+}$ is about $\rm \alpha_{O^{2+}}=1.4\times 10^{-11}cm^3 s^{-1}$. The above gas density is $n_{\rm H} = \rm 10^3 cm^{-3}$. According to the recombination time scale formula $\rm t_{rec }\sim 1/(\alpha n_{\rm H})$, the typical order of $\rm t_{rec}$ is $\rm 10^8 s$, i.e., a few years. This time scale is much smaller than the lifespan of quasars,
which is on the order of $10^7$ of years \citep{hopkins2005}. Compared to the lifespan of AGN, the hysteresis caused by the recombination time scale can be ignored.

\subsection{Possible influence of redshift and outflow velocity threshold}\label{sec:a4}
To investigate whether the relationship between the fraction of inflow and the AGN evolution is influenced by quasar redshift, we divided our sample into two groups based on the median redshift value ($z_{\rm median}=0.72$). Each group consisted of 1054 quasars. Within each group, we sorted the 1054 quasars based on the ratio $L_{\rm \oiii}$/$L_{3000}$ and divided them into 11 bins. The first ten bins contained 100 objects each, and the last bin contained 154 objects. The relationship between $F_{\rm inflow}$ and $L_{\rm \oiii}$/$L_{3000}$ for these 11 bins is displayed in Fig. \ref{fig:a5}. For the group of quasars with $z<0.72$, the best linear fit for $F_{\rm inflow}$ is given by $(-0.209\pm 0.031)\log_{10}(L_{\rm \oiii}$/$L_{3000})-(0.030\pm 0.076)$. The confidence level of the fitting slope, $\rho=-0.209\pm 0.031$, deviates significantly from the null hypothesis ($k=0$) at $6.8\sigma$. For the group of quasars with $z>0.72$, the best linear fit for $F_{\rm inflow}$ is given by $(-0.075\pm 0.033)\log_{10}(L_{\rm \oiii}$/$L_{3000})+(0.261\pm 0.082)$. The confidence level of the fitting slope, $\rho=-0.075\pm 0.033$, deviates from $k=0$ at $2.3\sigma$. The confidence level of $F_{\rm inflow}$ decreasing with increasing $L_{\rm \oiii}$/$L_{3000}$ is higher for the group of quasars with $z<0.72$ compared to the group of quasars with $z>0.72$. This difference may be attributed to the presence of more foreground-inserted non-intrinsic blue-shifted absorption lines (misidentified as outflows) in the group of quasars with $z>0.72$, which could interfere with the results. The fractions of outflows are 545/1054 and 586/1054 for the groups of quasars with $z<0.72$ and $z>0.72$, respectively. The higher fraction of blue-shifted absorption lines in the group of quasars with $z>0.72$ supports the aforementioned explanation.

To investigate the influence of threshold for outflow velocity on our result, we studied the correlation between the fraction of inflow and the $L_{\rm \oiii}$/$L_{3000}$ under two thresholds: 500 \kms\ (see panel (b) of Fig. \ref{fig2}), 750 \kms, and 1000 \kms. As the threshold increases, the correlation between the fraction of inflow or outflow and time evolution slightly decreases (see Fig. \ref{fig:a6}). The slope gradually flattens (index k from -0.146 to -0.119 for 500 \kms\ to 1000 \kms). This is a natural outcome. With the increase of the outflow velocity threshold, more central high-speed outflow is considered. The fraction of high-speed outflow from the center gradually decreases over time. Therefore, the mixing of high-speed outflow will lead to a decrease in the correlation between the fraction of galaxy scale outflow and time.

\addtocounter{figure}{-4}

\begin{figure*}
\renewcommand\thefigure{\textbf{a\arabic{figure}}}
\centering
\includegraphics[width=14.cm]{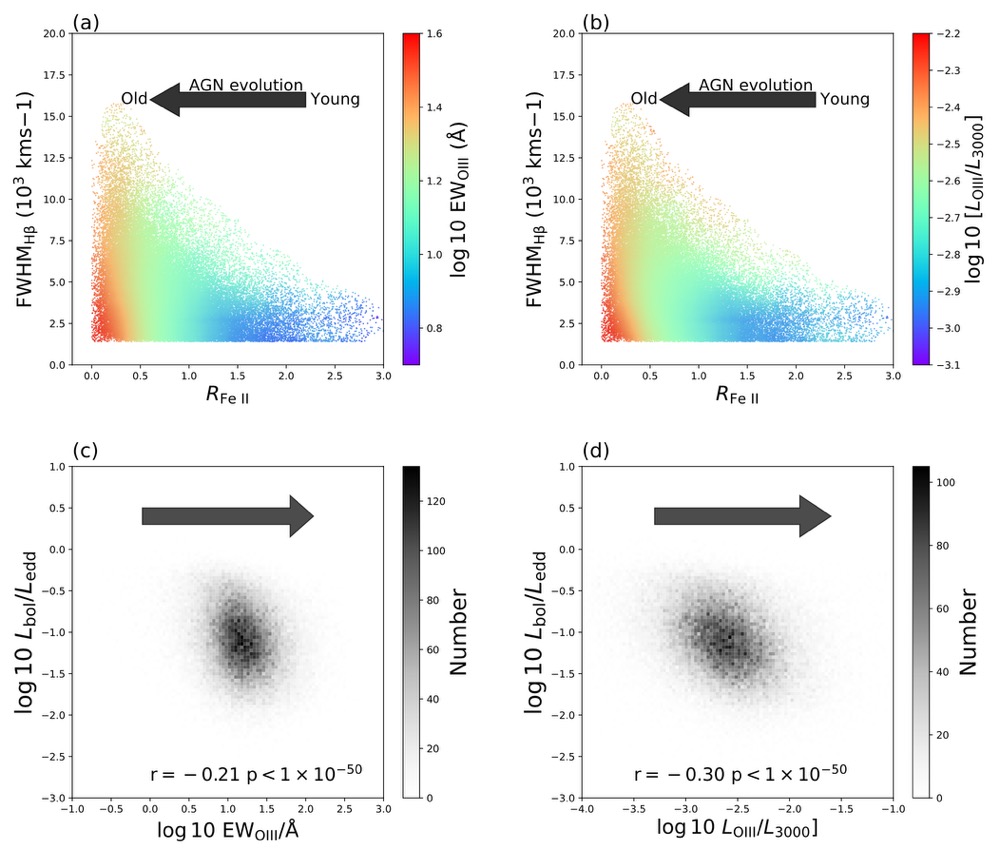}
\caption{Distribution of quasars in the EV1 plane for the SDSS DR16. Panels (a) and (b): the horizontal axis is the relative \feii\ strength, $R_{\rm \feii}$, and the vertical axis is the broad \hb\ FWHM. We colour-code the points by the \oiii\ strength averaged over all nearby objects in a smoothing box of $\Delta R_{\rm \feii}$=0.2 and $\Delta FWHM_{\hb} $=1,000 $\kms$. There is a systematic trend of decreasing \oiii\ strength with increasing $R_{\rm \feii}$. Previous studies suggested that the increasing \oiii\ strength could be considered as a tracer of the expansion of the ionization region as AGNs evolve (gray arrow). Panels (c) and (d): the inverse correlation between \oiii\ and $R_{\rm \feii}$ means that the evolution of AGN will be accompanied by a decrease in accretion rate. The Spearman correlation coefficient and corresponding uncorrelated probabilities are marked.}
\label{fig:a1}
\end{figure*}

\begin{figure*}
\renewcommand\thefigure{\textbf{a\arabic{figure}}}
\centering
\includegraphics[width=14.cm]{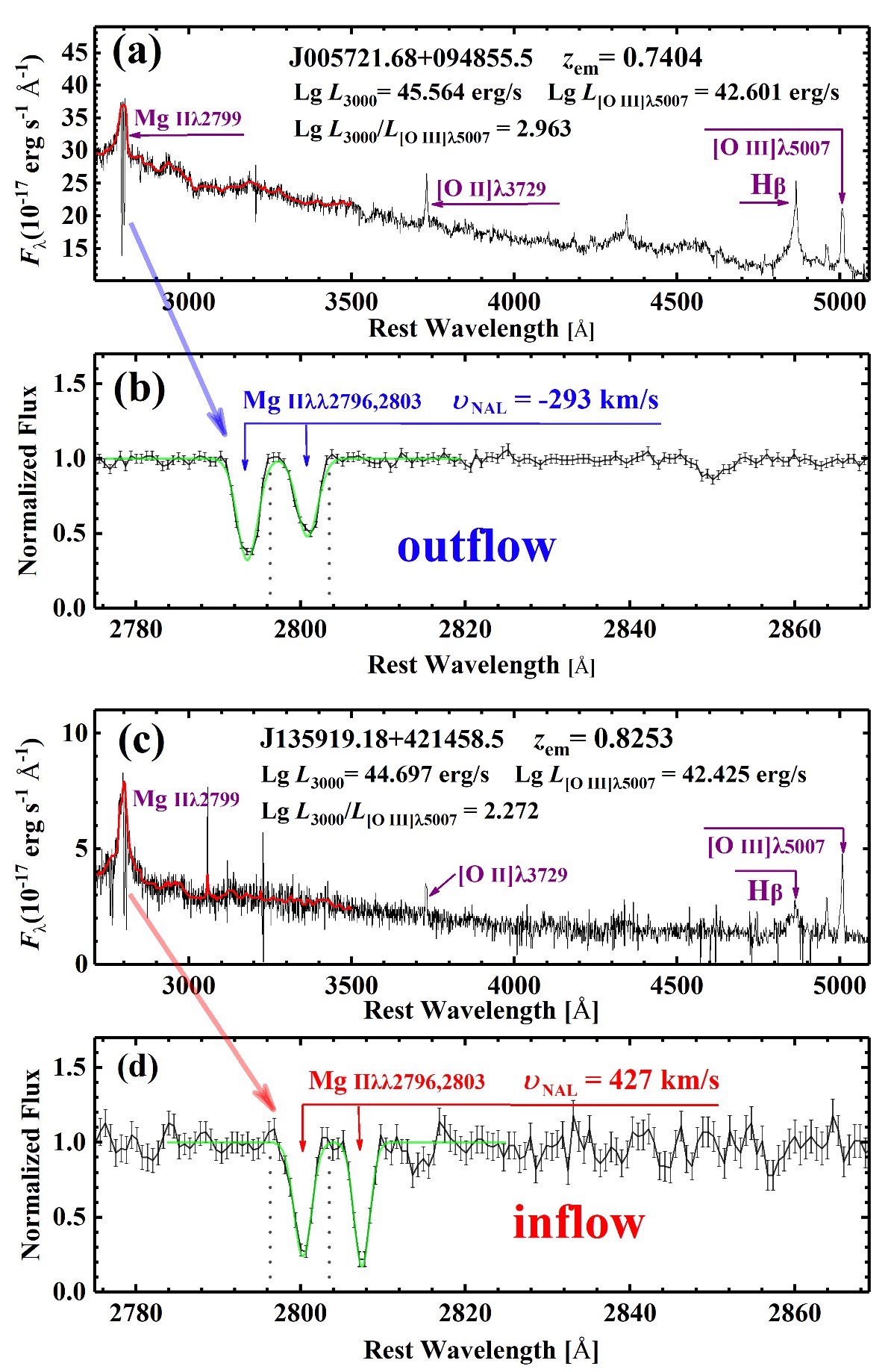}
\caption{Examples of the redshifted and blueshifted \mgii\ NALs. Panels (a) and (c) are the observed spectra. Panels (b) and (d) are the spectra normalized by corresponding pseudo-continuum fits around \mgii\ emission. Red solid-lines are the pseudo-continuum fits. Green solid-lines are the Gaussian function fits to the blueshifted or redshifted \mgii\ NALs. Black dot-lines label the positions of \Mgiiab\ NALs at quasar rest frame. }
\label{fig:a2}
\end{figure*}

\begin{figure*}
\renewcommand\thefigure{\textbf{a\arabic{figure}}}
\centering
\includegraphics[width=14.cm]{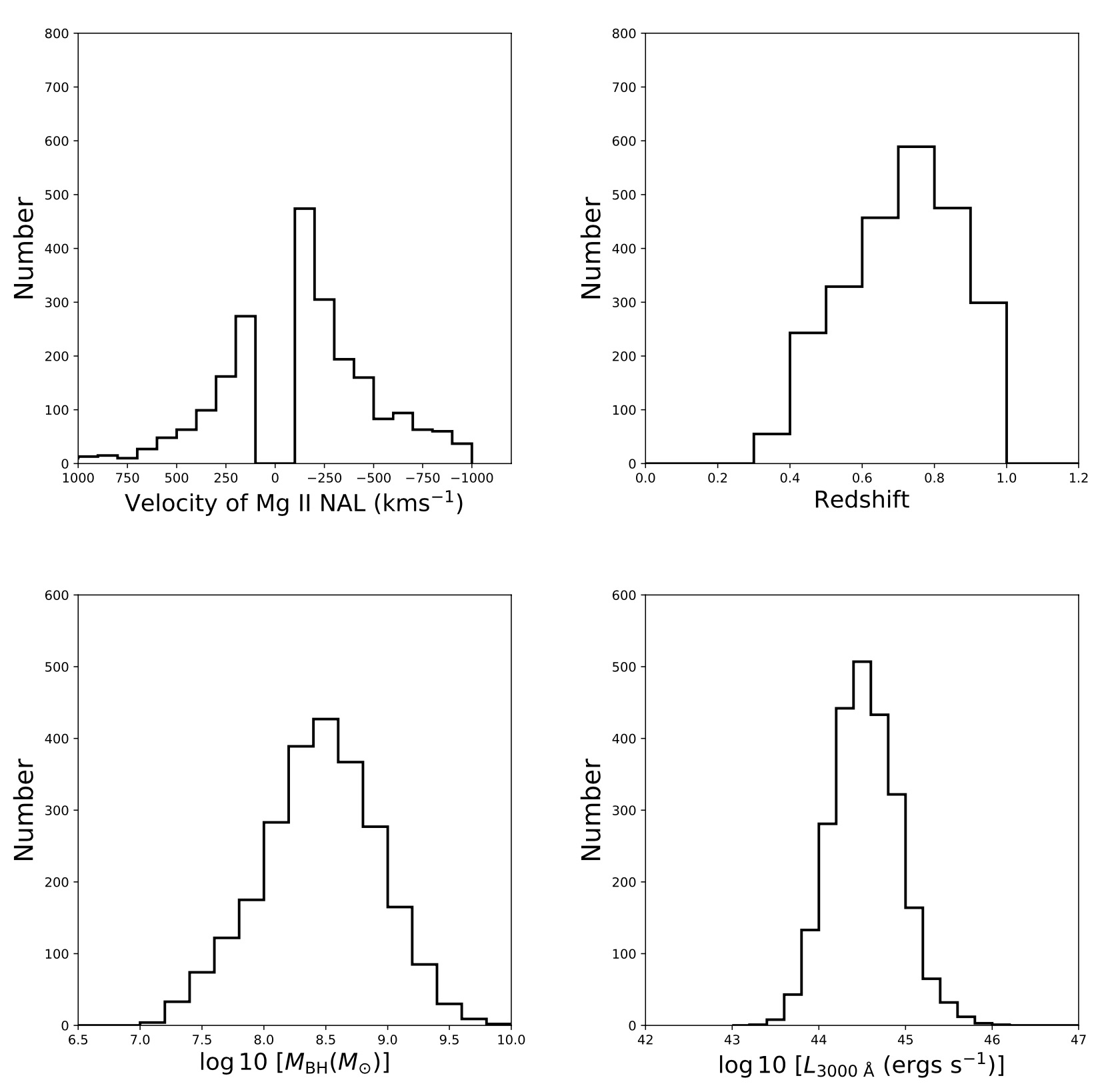}
\caption{The distributions of the velocities of \mgii\ NALs, redshifts of quasar systems, masses of SMBHs, and luminosities at 3000\AA~in our sample.}
\label{fig:a3}
\end{figure*}

\begin{figure*}
\renewcommand\thefigure{\textbf{a\arabic{figure}}}
\centering
\includegraphics[width=14.cm]{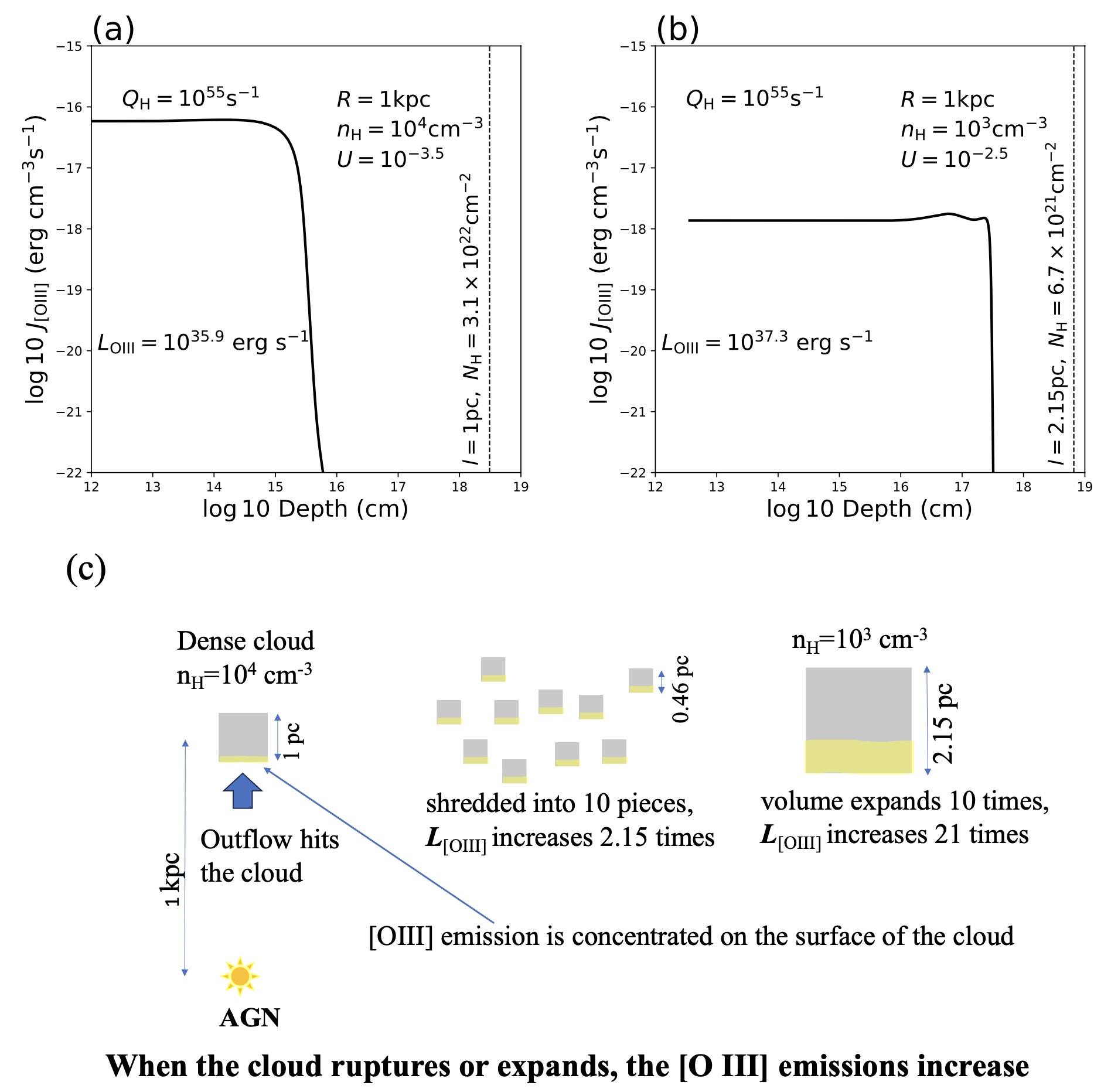}
\caption{Cloudy simulation for the \OIII\ emission. Panel (a): for a cube of gas cloud with volume density $n_{\rm H}=10^4\cc$, side length $l$ =1 pc and distance $R$=1 kpc from the central radiation source (with a typical Ionizing photon emissivity $Q_{\rm H}=10^{55}\rm s^{-1}$). The \oiii\ emission predominantly concentrates on the surface within depth of $d=10^{15.5}$ cm ($\sim$ 0.001 pc) and column density of  $N_{\rm H}=10^{19.5}\rm cm^{-2}$.
The total \oiii\ luminosity is $10^{35.9} \rm erg\ s^{-1}$. Panel (b): when the cloud expands 10 times ($n_{\rm H}$decreases from $10^4\cc$ to $10^3\cc$), the total \oiii\ luminosity increases to $10^{37.3} \rm erg\ s^{-1}$. Panel (c):  when the cloud cube is fragmented into 10 pieces, the \oiii\ emission increases 2.15 times due to the larger surface area. When the mass of cloud remains unchanged and the volume expands by 10 times, its \oiii\ emissions increase by 21 times. So, when the cloud ruptures or expands, the \oiii\ emissions increase.}\label{fig:a4}
\end{figure*}

\begin{figure*}
\renewcommand\thefigure{\textbf{a\arabic{figure}}}
\centering
\includegraphics[width=14.0cm]{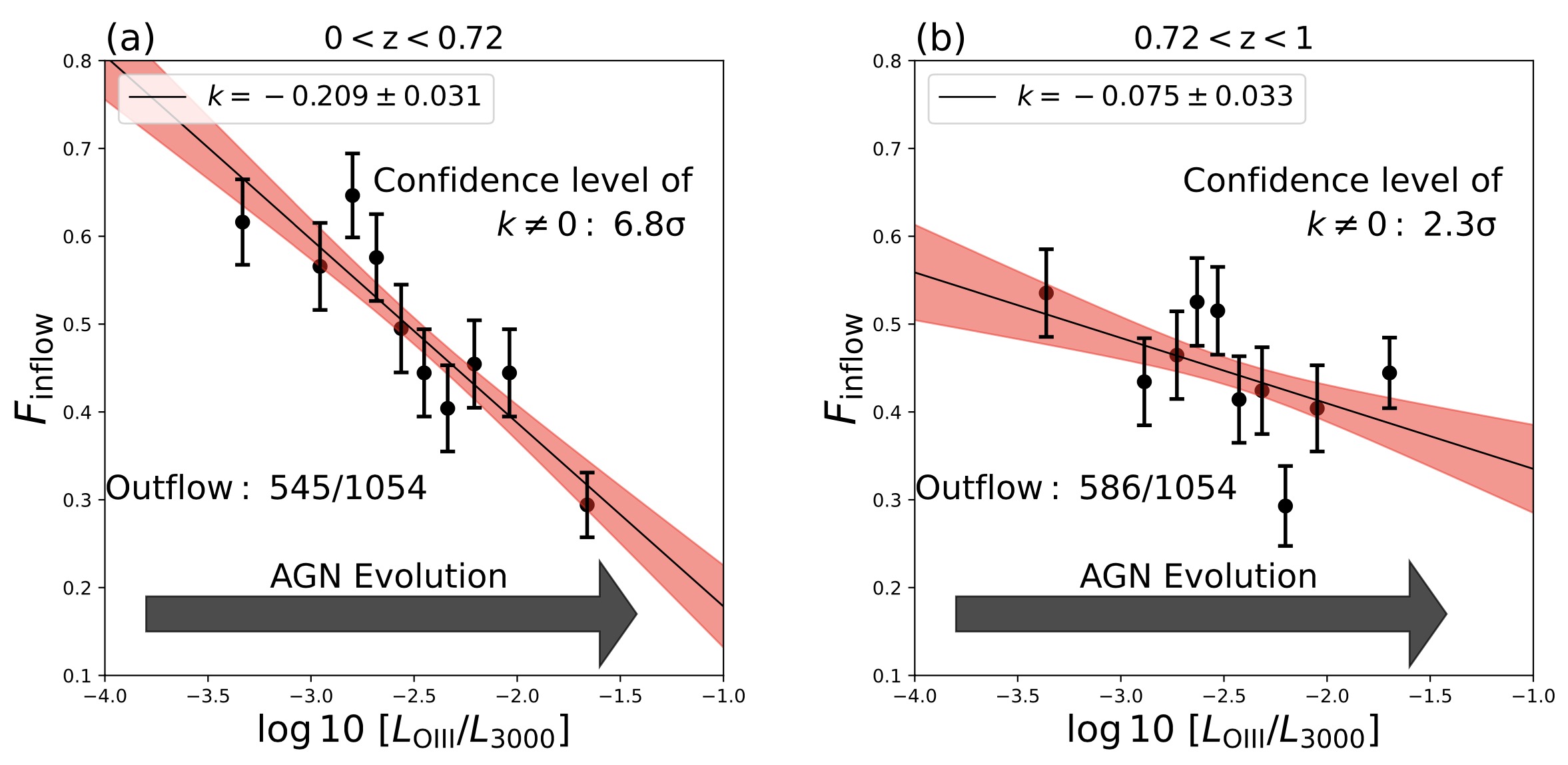}
\caption{The influence of quasar redshift on the transformation from inflow and outflow. We divide our sample into two parts according to the median value ($z_{\rm median}=0.72$) of quasar redshift. The slopes $k$ of the best linear fits and the confidence level of $k\neq 0$ are marked. The confidence level of $F_{\rm inflow}$ decreasing with the increase of $L_{\rm \oiii}$/$L_{3000}$ for $z<0.72$ is greater than that of $z>0.72$. The possible reason is that there are more foreground inserted non intrinsic blue-shifted absorption lines in group of $z>0.72$ that interfere with the results. The higher fraction of blue shift absorption lines in group of $z>0.72$ also supports the above reason. }
\label{fig:a5}
\end{figure*}

\begin{figure*}
\renewcommand\thefigure{\textbf{a\arabic{figure}}}
\centering
\includegraphics[width=14.0cm]{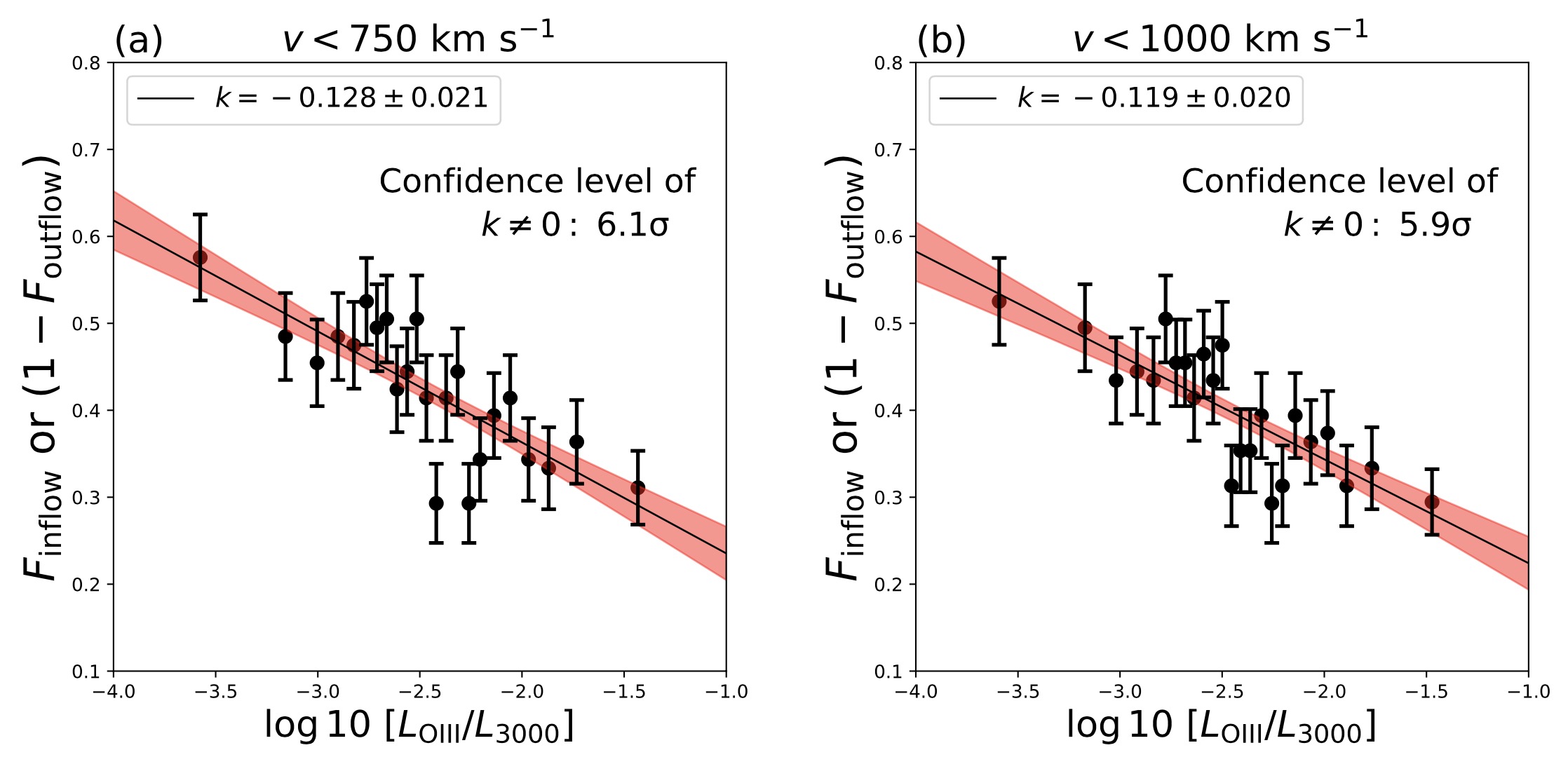}
\caption{The influence of threshold for outflow velocity: 500\kms (see panel (b) of Fig. \ref{fig2}), 750 \kms, and 1000 \kms. As the threshold increases, the correlation
between the fraction of inflow or outflow and time evolution slightly decreases. This is a natural outcome. With the increase of the outflow velocity threshold, more central high-speed outflow is considered. The fraction of high-speed outflow from the center gradually decreases over time. Therefore, the mixing of high-speed outflow will lead to a decrease in the correlation between the fraction of galaxy scale outflow and time.}
\label{fig:a6}
\end{figure*}


\end{document}